\documentstyle[12pt]{article}

\newcommand{\be}{\begin{equation}}
\newcommand{\ee}{\end{equation}}
\newcommand{\ba}{\begin{eqnarray}}
\newcommand{\ea}{\end{eqnarray}}
\begin{document}
\def\pct#1{\input epsf \centerline{ \epsfbox{#1.eps}}}

\begin{titlepage}
\hbox{\hskip 12cm ROM2F-95/09  \hfil}
\hbox{\hskip 12cm CPTh/RR.358.0695 \hfil}
\hbox{\hskip 12cm June \ 1995 \hfil}
\vskip 1.5cm
\begin{center}
{\Large  \bf  The \ Open \ Descendants \ of \\
\vskip .8cm
Non-Diagonal \ $SU(2)$ \ WZW \ Models}

\vspace{1.8cm}

{\large G. Pradisi${}^{a,b,c}$ , \ A. Sagnotti${}^{a,b}$ \ \ and \ \  Ya.S.
Stanev${}^{a,b,}$\footnote{I.N.F.N.  Fellow, on
Leave from Institute for Nuclear Research and Nuclear Energy, Bulgarian
Academy of Sciences, BG-1784 Sofia, BULGARIA.}}

\vspace{0.8cm}

${}^{a}${\sl Dipartimento di Fisica\\
Universit{\`a} di Roma \ ``Tor Vergata'' \\
I.N.F.N.\ - \ Sezione di Roma \ ``Tor Vergata'' \\
Via della Ricerca Scientifica, 1 \ \
00133 \ Roma \ \ ITALY}
\vskip 14pt
${}^{b}${\sl Centre de Physique Th{\'e}orique \\
Ecole Polytechnique \\
91128 Palaiseau \ \ FRANCE}
\vskip 14pt
${}^{c}${\sl Centro ``Vito Volterra'', Universit{\`a} di Roma
``Tor Vergata''}
\vspace{0.5cm}
\end{center}
\vskip 0.8cm
\abstract{We extend the construction of open descendants to the
$SU(2)$ WZW models with non-diagonal left-right pairing, namely $E_7$ and the
$D_{odd}$ series in the $ADE$ classification of Cappelli, Itzykson
and Zuber.  The structure of the resulting models is
determined to a large extent by the ``crosscap constraint'', while their
Chan-Paton
charge sectors may be embedded in a general fashion into those of the
corresponding
diagonal models.
}
 \vfill
\end{titlepage}
\addtolength{\baselineskip}{0.3\baselineskip}

\vskip 24pt
\begin{flushleft}
{\large \bf Introduction}
\end{flushleft}

We have recently shown \cite{pss} how ``open descendants'' \cite{bs}
can be associated
to all diagonal $SU(2)$ WZW models \cite{wzw}. In this letter we extend
the program of ref. \cite{car} to the remaining, non-diagonal, ones,
thus completing the classification for all the $ADE$ modular
invariants of Cappelli, Itzykson and Zuber \cite{ciz}. This is particularly
instructive, since the ansatz suggested by ref. \cite{cardy}
is tailored to the diagonal case.
The extension of ref. \cite{bps} to the non-diagonal minimal models of
the $ADE$ classification \cite{ciz} will be discussed elsewhere.
We also sketch a general method to handle the constraints
and exhibit the relationship between these models and the
corresponding diagonal
ones. The transverse-channel amplitudes are determined to
a large extent by the
crosscap constraint of ref. \cite{fps}, and the resulting open descendants
are somewhat surprising, since their Chan-Paton \cite{cp} charge
spaces are typically far smaller than one would naively expect.

Let us begin by reviewing the general features of the construction
of open descendants.
The starting point is a left-right symmetric two-dimensional
conformal field theory whose
spectrum is described by a torus partition function
\be
T \ = \ \sum_{a,b} \  \chi_a ( \tau ) \ N_{ab} \ {\bar \chi}_b ( {\bar \tau} )
\label{torus}
\ee
invariant under the modular group $PSL(2,Z)$.  As is well known, this group is
generated by the transformations
\be
T: \ \tau \rightarrow \tau + 1 \qquad \quad {\rm and } \quad \qquad
S: \ \tau \rightarrow - \ {1 \over \tau} \quad ,
\label{TandS}
\ee
that act on the characters $ \chi_a $ via two matrices, also
denoted by $T$ and
$S$.  In moving to the open descendants, one begins by projecting
the closed spectrum with a direct-channel Klein-bottle amplitude
$K$, that has the general form
\be
 K \ = \ {1 \over 2} \ \sum_a \ \chi_a \ K_a \qquad {{ S \atop \longrightarrow}
\atop } \qquad
\tilde K \ = \ {1 \over 2} \ \sum_a \ \chi_a \left( \Gamma_a \right)^2
\label{klein}
\ee
and, as indicated, is turned into the corresponding transverse-channel
amplitude $\tilde K$
by an $S$ transformation.  In  addition, the descendants contain ``open''
sectors, described by direct-channel annulus and M{\"o}bius
amplitudes, that must respect the planar duality and the
factorization of disk amplitudes. The direct-channel annulus amplitude
$A$ has the general form
\be
 A \ = \ {1 \over 2} \ \sum_{a, \beta , \gamma} \ \chi_a \ A_{a \beta \gamma} \
n_{\beta} \
n_{\gamma} \qquad
{ {S \atop \longrightarrow} \atop } \qquad
\tilde A \ = \ {1 \over 2} \ \sum_a \ \chi_a \left( \sum_{\beta} \  B_{a \beta}
\
n_{\beta} \right)^2
\label{annulus}
\ee
and, as indicated, is turned
into the transverse-channel amplitude $\tilde A$ by an $S$ transformation.
Finally,
the direct-channel M{\"o}bius amplitude $M$ has the general form
\be
 M \ = \ \pm \ { 1 \over 2} \ \sum_{a, \beta} \ {\hat \chi_a} \ M_{a \beta} \
n_{\beta}  \qquad
{{ P \atop \longrightarrow}\atop } \qquad
\tilde M \ = \ \pm \ { 1 \over 2} \ \sum_a \ {\hat \chi}_a \
\Gamma_a \  \left( \sum_{\beta} \ B_{a \beta}
\ n_{\beta} \right) \quad ,
\label{moebius}
\ee
and is turned into the corresponding transverse-channel amplitude $\tilde M$ by
the transformation $P:{i \tau  + 1 \over 2} \rightarrow
{i   + \tau \over 2 \tau}$, that acts on the characters via the matrix
\be
P=T^{1/2} \ S \ T^2 \ S \ T^{1/2} \qquad .
\label{pmat}
\ee
The three matrices $S$, $T$ and $P$ satisfy the relations
\be
S^2 \ = \ ( S T )^3 \ = \ P^2 \ = \ C \qquad , \label{modular}
\ee
where $C$ is the conjugation matrix.
The resulting models extend 2D Conformal Field Theory to the open and
unorientable case
and are building blocks for the construction of open-string vacua.
The three transverse-channel
amplitudes $\tilde K$, $\tilde A$ and $\tilde M$  are closely related, since
they
describe the propagation of the closed spectrum along tubes
terminating at holes and/or crosscaps.  As a result, each coefficient in
$\tilde M$ is the geometric mean of the corresponding ones in
$\tilde A$ and in $\tilde K$.
Moreover, the integer coefficients $K_a$,
$M_{a \beta}$ and $A_{a \beta \gamma}$ describe consistently
the (anti)symmetrization of the
spectrum both in the closed and in the open sectors,
provided they satisfy the relations
\be
| K_a | \ = \ N_{aa} \qquad , \qquad  M_{a \beta} \ = \ A_{a \beta \beta}
\quad {\rm \ ( mod \ 2 )} \qquad .
\label{constraints}
\ee
As discussed in ref. \cite{pss}, this form of the last condition allows
for multiple charge sectors corresponding
to a single character. These constraints lead in general
to rather complicated systems of diophantine equations for the tensors
$K_a$, $M_{a \beta}$ and $A_{a \beta \gamma}$, whose solution determines the
Chan-Paton charge space of the open sector, as well as the corresponding
projection of the closed sector.

\vskip 24pt
\begin{flushleft}
{\large \bf Descendants of the $D_5$ Model}
\end{flushleft}

Let us  illustrate the construction on the simplest example, the
 $D_5$ model of level $k=6$,
whose torus partition function is
\be
T \ = \ |\chi_1|^2 \ + \ |\chi_3|^2 \ + \ |\chi_5|^2 \ + \ |\chi_7|^2 \ + \
\ |\chi_4|^2 \ + \left( \chi_2 {\bar \chi}_6 + h.c. \right) \label{tord3}
\ee
where the subscript of $\chi_a$ is related to the isospin $I$ by
$a = 2I+1$.  This model admits two classes
of open descendants corresponding to the two Klein-bottle projections
\be
K^r \ = \ {1 \over 2} \ \left(  \chi_1 \ + \ \chi_3 \ + \ \chi_5 \ + \ \chi_7 \
- \
\ \chi_4 \right) \label{kld3r}
\ee
and
\be
K^c \ = \ {1 \over 2} \ \left(  \chi_1 \ + \ \chi_3 \ + \ \chi_5 \ + \ \chi_7 \
+ \
\ \chi_4 \right) \qquad .\label{kld3c}
\ee
In both cases only the
five self-conjugate
characters $\chi_1$, $\chi_3$, $\chi_5$, $\chi_7$ and $\chi_4$
are allowed in the transverse-channel annulus
amplitude.  Thus, in analogy with more elementary examples,
both classes of models would be
expected to contain five charge sectors. Rather surprisingly,
however, they only admit four
charge sectors.  Indeed, turning on one charge at a time
effectively splits the constraints into independent ones that
link the {\it diagonal} elements in the annulus amplitude
to the corresponding
entries in the M{\"o}bius amplitude. Eqs. (\ref{klein}),
(\ref{annulus}), (\ref{moebius}) and (\ref{constraints}) may then
be solved,
and the consistent solutions may be combined into the general
parametrization. The annulus and M{\"o}bius amplitudes are then
\ba
\lefteqn{A^r \ = \ {1 \over 2} \ \biggl(
\chi_1 ( l_1^2 + l_2^2 + l_3^2 + l_4^2 ) \ + \
(\chi_2 + \chi_6 )( 2 l_1 l_2 ) \ +} \nonumber \\
& & \chi_3 ( l_1^2 + 2 l_1 l_3 + 2 l_1 l_4 + 2 l_3 l_4 ) \ + \ \chi_4 ( 2 l_2
l_3 +
2 l_2 l_4 )  \ + \nonumber \\
& &\chi_5 (l_1^2  + l_3^2  + l_4^2 + 2 l_1 l_3 + 2 l_1 l_4 ) \ + \
\chi_7 ( l_1^2 + l_2^2 + 2 l_3 l_4 ) \biggr) \quad , \label{ad3r}
\ea
and
\ba
M^r &=& \pm \ {1 \over 2} \ \biggl( {\hat \chi}_1 ( l_1 - l_2 + l_3 +
l_4 ) \ + \ {\hat \chi}_3 ( - l_1) \ +
\nonumber \\
& & \qquad {\hat \chi}_5 (l_1 + l_3 + l_4 ) \ +
\ {\hat \chi}_7 ( l_1 + l_2 ) \biggr) \label{md3r}
\ea
for the model with all real charges, and
\ba
\lefteqn{A^c \ = \ {1 \over 2} \ \biggl( \chi_1 ( l_1^2 + l_2^2 + 2 l {\bar l}
) \ + \
(\chi_2 + \chi_6 )( 2 l_1 l_2 ) \ +} \nonumber \\
& & \chi_3 (l_1^2  + l^2  + {\bar l}^2 + 2 l_1 l + 2 l_1 {\bar l} ) \ +
\ \chi_4 ( 2 l_2 l + 2 l_2 {\bar l} )  \ + \nonumber \\
& & \chi_5 ( l_1^2 + 2 l_1 l + 2 l_1 {\bar l} + 2 l {\bar l} ) \ + \
\chi_7 ( l_1^2 + l_2^2 + l^2 + {\bar l}^2 ) \biggr)  \label{ad3c}
\ea
and
\ba
M^c &=& \pm \ {1 \over 2} \ \biggl( {\hat \chi}_1 ( - l_1 + l_2 ) \ + \
{\hat \chi}_3 (l_1 + l +
{\bar l} ) \ + \nonumber \\
& & \qquad {\hat \chi}_5 ( l_1 ) \ + \
{\hat \chi}_7 ( l_1 + l_2 + l + {\bar l} ) \biggr) \label{md3c}
\ea
for the model with complex charges.  It should be appreciated
that in the latter model only the two real charges $l_3$ and $l_{4}$
turn into the complex pair $(l,{\bar l})$ where, as usual, $l=\bar{l}$.
This is in sharp contrast with the behavior of the complex
diagonal models, that contain  at most one real charge.
Although these amplitudes are not
directly based on the fusion rules, an explicit study of
the disk amplitudes
shows that the open spectrum is nicely consistent
with planar duality and factorization.
\vskip 24pt
\begin{flushleft}
{\large \bf The Crosscap Constraint}
\end{flushleft}

In this Section we provide an alternative, rigorous derivation
of the crosscap constraint of ref. \cite{fps}.
Let us begin by noting that, in left-right symmetric models, all fields
of the closed sector may be expressed in terms of chiral and antichiral
vertex operators \cite{tk} according to
\be
\phi_{\Delta , \bar{\Delta}}(z,\bar{z}) \ = \ \sum_{i, \bar{i}, f, \bar{f}} \
{{V_{\Delta}}^f}_i (z) \
{{{\bar V}_{\bar \Delta} \, }^{\bar f}}_{\bar i} (\bar z) \
{\alpha_{i \bar i}}^{f \bar{f}} \qquad , \label{chiral1}
\ee
where ${{V_{\Delta}}^f}_i$ denotes a
chiral ``field'' of conformal dimension $\Delta$ acting on a state $i$ and
producing a state $f$ and $\bar V$ denotes a corresponding
antichiral ``field''.  In order to analyze the behavior of the fields
$\phi_{\Delta , \bar{\Delta}}(z,\bar{z})$ in the
presence of a crosscap, it is convenient to introduce also
\be
\phi_{\bar{\Delta},\Delta}(\bar{z},z) \ = \ \sum_{i, \bar{i}, f, \bar{f}} \
{{V_{\bar{\Delta}}}^{\bar{f}}}_{\bar{i}} (\bar{z}) \ {{\bar{V}_{\Delta} \,
}^f}_i (z) \
{\alpha_{i \bar{i}}}^{f \bar{f}} \qquad . \label{chiral2}
\ee
For {\it all} fields in the diagonal ($A$-series) models $\Delta$ and
$\bar{\Delta}$ coincide,
and ${\alpha_{i \bar{i}}}^{f \bar{f}} = \delta_{i \bar{i}} \
\delta^{f \bar{f}}$, while for the $D_{odd}$ models with $k=2 \ {\rm mod} \ 4$,
$\bar{\Delta} = \sigma(\Delta)$,
${\alpha_{i \bar{i}}}^{f \bar{f}} = \delta_{i \sigma(\bar{i})} \
\delta^{f \sigma(\bar{f})}$, where $\sigma(I)= I$ for $I$ integer and
$\sigma(I)= k/2-I$ for $I$ half-odd integer.

The $n$-point functions in front of a crosscap can be obtained from the
$2n$-point chiral conformal blocks, since the crosscap interchanges
left and right chiral vertex operators with the same labels.  More precisely,
one can
account for the crosscap by inserting in the amplitudes a ``crosscap operator''
that
acts according to
\be
{\hat C} \ = \ \sum_l \ \Gamma_l \ | \Delta_l > < {\bar \Delta}_l | \qquad ,
\label{crosscap}
\ee
where, as in ref. \cite{fps}, $\Gamma_l$ is the normalization of the one-point
function of a primary field of conformal dimensions $(\Delta_l, {\bar \Delta}_l
)$
in front of the crosscap. The coefficients $\Gamma_l$ are a crucial ingredient
of the
construction,
since their squares determine the vacuum Klein-bottle amplitude. The $n$-point
functions in front of a crosscap are then defined as
\be
{< \phi_{\Delta_1, {\bar \Delta}_1} \ .. \ \phi_{\Delta_n, {\bar \Delta}_n}
>}_C
\ = \ {< {\hat C} \phi_{\Delta_1, {\bar \Delta}_1} \ .. \ \phi_{\Delta_n, {\bar
\Delta}_n} >}_0
\quad , \label{npoint1}
\ee
and can be expressed in terms of the chiral conformal blocks for $2n$
chiral vertex operators.
Indeed, using eqs. (\ref{chiral1}) and (\ref{crosscap}),
\ba
{< \phi_{\Delta_1, {\bar \Delta}_1} .. \phi_{\Delta_n, {\bar \Delta}_n} >}_C
 &=& \sum_l  \Gamma_l  < 0 | V_{\Delta_1}(z_1) .. V_{\Delta_n}(z_n) |
\Delta_l  >
 < {\bar \Delta}_l | {\bar V}_{{\bar \Delta}_1}({\bar z}_1) ..
{\bar V}_{{\bar \Delta}_n}({\bar z}_n) | 0 > \nonumber \\
&=&  \sum_l  \Gamma_l  {< 0 | V_{\Delta_1}(z_1) .. V_{\Delta_n}(z_n)  \
V_{{\bar \Delta}_1}({\bar z}_1) .. V_{{\bar \Delta}_n}({\bar z}_n) | 0 >}_l
\quad .
\label{unwrap}
\ea
The (anti)symmetrization properties of the fields $\phi_{\Delta_i, {\bar
\Delta}_i}$
with respect to the crosscap then imply the relation
\be
{< \phi_{\Delta_i, {\bar \Delta}_i} (z_i, {\bar z}_i ) \ X >}_C \ = \
\varepsilon_{( i, \bar i )} \
{< \phi_{{\bar \Delta}_i, \Delta_i} ({\bar z}_i, z_i ) \ X >}_C \quad ,
\label{phase}
\ee
where $X$ is any polynomial in the fields.  It should be noticed that
$\varepsilon$ coincides with the sign in the Klein-bottle projection
(\ref{klein}) only for the integer-isospin fields, while it is opposite for the
half-odd-integer ones, as befits the pseudoreality of the half-odd-integer
isospin representations
of $SU(2)$. Eq. (\ref{phase})  generally determines all the coefficients
$\Gamma_l$.
In particular, for all fields with
$\varepsilon =-1$, as well as for those of non-zero spin ({\it i.e.} with
$\Delta \ne {\bar \Delta}$), $\Gamma$ must vanish, since
\ba
 { < \phi_{\Delta, {\bar \Delta}} (z, {\bar z} )  >}_C \
&=& \sum_l  \Gamma_l  < 0 | V_{\Delta}(z) | \Delta_l >
 < {\bar \Delta}_l | {\bar V}_{\bar \Delta}({\bar z}) | 0 > \nonumber \\
&=& \Gamma_{\Delta} \ \delta_{\Delta, {\bar \Delta}} \ < 0 | V_{\Delta}(z) \
V_{\bar \Delta}({\bar z}) | 0 >\ = \ {< \phi_{{\bar \Delta}, \Delta} ({\bar
z},z )  >}_C
\quad . \label{onepoint}
\ea

Much more detailed information can be obtained by analyzing the two-point
functions
in front of a crosscap
\ba
\lefteqn{< \phi_{\Delta_1, {\bar \Delta}_1} (z_1, {\bar z}_1 )  \
 \phi_{\Delta_2, {\bar \Delta}_2} (z_2, {\bar z}_2 )  >_C } \hspace{3.5cm}
\nonumber \\
 =& & \sum_l  \Gamma_l  < 0 | V_{\Delta_1}(z_1) \ V_{\Delta_2}(z_2) | \Delta_l
>
 < {\bar \Delta}_l | {\bar V}_{{\bar \Delta}_1}({\bar z}_1) \
{\bar V}_{{\bar \Delta}_2}({\bar z}_2) | 0 > \nonumber \\
=& &  \sum_l  \Gamma_l  < 0 | V_{\Delta_1}(z_1) \ V_{\Delta_2}(z_2)  \
V_{{\bar \Delta}_1}({\bar z}_1) \  V_{{\bar \Delta}_2}({\bar z}_2) | 0 >_l
\nonumber \\
=& &  \sum_l  \Gamma_l  \ {C_{( \Delta_1, {\bar \Delta}_1 )( \Delta_2, {\bar
\Delta}_2 )}}
^{( \Delta_l, \Delta_l )} \ S_{l}(z_1, z_2,{\bar z}_1,{\bar z}_2 ) \quad ,
\label{twopoint1}
\ea
where $S_l$ denote the normalized $s$-channel conformal blocks with a field of
dimensions $( \Delta_l, \Delta_l )$ in the intermediate channel, while
${C_{( \Delta_1, {\bar \Delta}_1 )( \Delta_2, {\bar \Delta}_2 )}}
^{( \Delta_l, \Delta_l )}$ are the two-dimensional structure constants.  Note
that we have set ${\bar \Delta}_l = \Delta_l$, since only spin-zero fields are
allowed
as intermediate states and, for brevity, we have omitted the additional labels
of the
chiral vertex operators.  In a similar fashion, one obtains
\ba
\lefteqn{< \phi_{{\bar \Delta}_1 , \Delta_1} ({\bar z}_1 , z_1)  \
 \phi_{\Delta_2, {\bar \Delta}_2} (z_2, {\bar z}_2 )  >_C }  \hspace{3.5cm}
\nonumber \\
=& &
\sum_l  \Gamma_l  < 0 | V_{{\bar \Delta}_1}({\bar z}_1) \ V_{\Delta_2}(z_2) |
\Delta_l >
 < {\bar \Delta}_l | {\bar V}_{{\Delta}_1}( z_1) \
{\bar V}_{{\bar \Delta}_2}({\bar z}_2) | 0 > \nonumber \\
=& & \sum_l  \Gamma_l  < 0 | V_{{\bar \Delta}_1}({\bar z}_1) \
V_{\Delta_2}(z_2)  \
V_{{\Delta}_1}( z_1) \  V_{{\bar \Delta}_2}({\bar z}_2) | 0 > \nonumber \\
=& & \sum_l  \Gamma_l  \ {C_{( {\bar \Delta}_1 , \Delta_1 )( \Delta_2, {\bar
\Delta}_2 )}}
^{( \Delta_l, \Delta_l )} \ S_{l}({\bar z}_1, z_2, z_1,{\bar z}_2 ) \quad .
\label{twopoint2}
\ea

The $s$-channel blocks in eqs. (\ref{twopoint1}) and (\ref{twopoint2}) can be
related
with the help of the duality matrices, as shown in figure 1.
\vskip 15pt
\input epsf \centerline{ \epsfbox{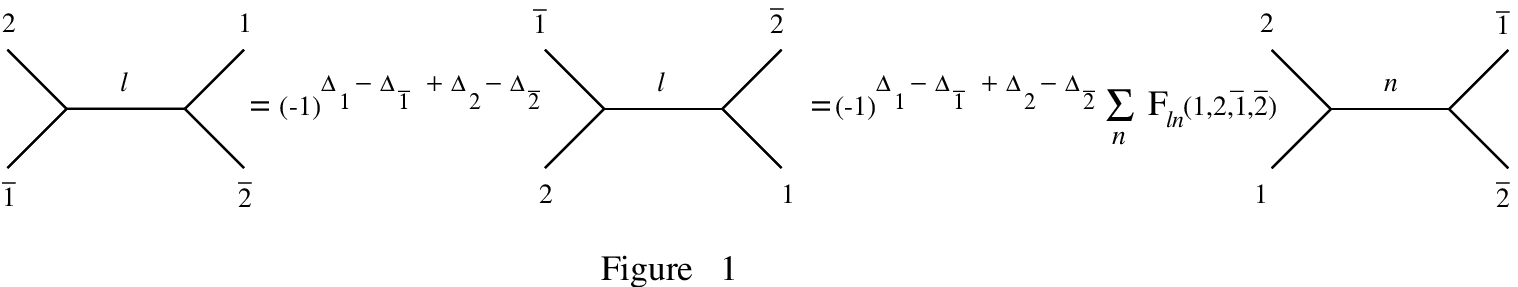}}
\vskip 15pt
\noindent
The first equality in the figure holds since the corresponding
transformation is $B_1 ({B_3})^{-1}$, where $B_i$ are the exchange matrices for
the conformal blocks \cite{ms}, while the second one follows from the
definition
of the fusion matrix $F$.  Therefore
\be
S_{l}({\bar z}_1, z_2, z_1,{\bar z}_2 ) \ = \ {( - 1 )}^{{\Delta}_1 - {\bar
\Delta}_1
+ {\Delta}_2 - {\bar \Delta}_2} \ \sum_n \ F_{ln} ( 1,2,{\bar 1},{\bar 2} ) \
S_{n}(z_1, z_2, {\bar z}_1,{\bar z}_2 ) \label{sduality} \quad ,
\ee
and inserting this expression in eqs.
(\ref{twopoint1}) and (\ref{twopoint2}) and using
eq. (\ref{phase}) yields a set of linear relations between the
crosscap coefficients $\Gamma_l$,
\ba
\lefteqn{\varepsilon_{( 1 ,{\bar 1} )} \ {( - 1 )}^{{\Delta}_1 - {\bar
\Delta}_1
+ {\Delta}_2 - {\bar \Delta}_2} \ \Gamma_n \ {C_{( \Delta_1, {\bar \Delta}_1 )(
\Delta_2, {\bar \Delta}_2 )}}
^{( \Delta_n, \Delta_n )} \ =} \hspace{3.5cm} \nonumber \\
 & &\sum_l \ \Gamma_l  \ {C_{( {\bar \Delta}_1 , \Delta_1 )( \Delta_2, {\bar
\Delta}_2 )}}
^{( \Delta_l, \Delta_l )} F_{ln} ( 1,2,{\bar 1},{\bar 2} ) \quad .
\label{cconstraint}
\ea
This ``crosscap constraint'' was first derived in ref. \cite{fps} where,
however,
only the $\varepsilon = 1$ case was considered.

The crosscap constraint can be looked at from two different viewpoints.  On the
one
hand, if the structure constants $C$ are known, eq. (\ref{cconstraint}) can be
used
to determine the values of
the one-point functions $\Gamma_l$.  On the other hand, if the one-point
functions are
(even partially) known, for instance from the vacuum Klein-bottle amplitude,
eq. (\ref{cconstraint}) can be used to determine the structure constants in the
model.
In particular, it allows one to determine rather simply the relative signs of
the
structure constants in the diagonal and non-diagonal models.

Since the $F$ matrix depends only on the chiral content of the
theory, it is the same for the diagonal $A$ models and for the non-diagonal
$D_{odd}$ models  of the same level $k$.  If, without any loss of generality,
one orders the
isospins of the fields in eq. (\ref{cconstraint}) so that they satisfy
\be
I_4 = {\rm min} \{ I_i \} \ , \quad | I_{12} | \le I_{34}  \ , \quad | I_{23} |
\le I_{14}
\ , \quad {\rm where} \quad I_{ij} = I_i - I_j \qquad ,
\label{restrictions}
\ee
$F$ has the form \cite{rst}
\ba
& &F_{I_{14}+r,I_{34}+s}(1,2,3,4) = \ \sum_{j=0}^{{\rm min} (s,m-r)} \ (-1)^{m
- r - j} \
 {{ [ 2 I_{34} + s + j ]! [ I_{34} - I_{12} + s ]!  } \over
{ [ 2 I_{34} + 2 s ]! [ I_{34} - I_{12} + j ]!  }} \times
\nonumber \\
& & \quad  \qquad {{ [ s ]!
[ m - j ]! [ m - j - I_{32} + I_{14} ]! [ 2 I_{14} + 2 r + 1 ]!
} \over {
 [ s - j ]! [ j ]!
[ m - r - j ]! [ r ]! [ r - I_{32} + I_{14} ]! [ 2 I_{14} + m + r - j +1 ]!
}} \ \; ,
\label{dualityf}
\ea
where
$m=2 I_4$ and
\ba
[ n ] &=& {{q^n - q^{-n}} \over {q - q^{-1}}} \quad , \qquad \qquad
q = exp( {{i \pi} \over {k+2}} )
\quad , \nonumber \\ {[ n ]!} &=& [ n ] [n-1]! \quad , \quad \qquad
[0]! = [1]! = 1 \quad . \label{quantum}
\ea
Moreover,
\ba
& & {C_{(1,3)(2,4)}}^{(I,I)} \ = \ \epsilon_{I} \ \sqrt{C_{12I} C_{34I}
 \over C_{II0}}
\quad , \\ \nonumber
\\
& & C_{12I} = {{[I_1+I_2+I+1]! [I_1+I_2-I]! [I_1+I-I_2]! [I_2+I-I_1]!} \over
{[2I_1]! [2I_2]! [2I]!}} \quad ,
\label{chiralc}
\ea
and $\epsilon_I = +1$ for the diagonal $(A)$ models, while for the non-diagonal
$(D_{odd})$ models
\ba
\epsilon_I \ &=& \ 1 \qquad \ {\rm for \ \ I_1,...,I_4 \ \ integers} \nonumber
\\
\epsilon_I \ &=& \ (-1)^I \qquad {\rm for \ \ I_1,...,I_4 \ \
half-odd-integers} \quad . \label{signs}
\ea
This exhausts all the cases we need, since in the $D_{odd}$ models the
one-point functions
$\Gamma$ vanish for all fields of half-odd-integer isospin.
Hence $I$ in eq. (\ref{signs})
is always an integer, while $I_1$ and $I_3$ are either both integer or both
half-odd-integer,
and the same is true for $I_2$ and $I_4$. It should be noticed that
both $F$ and the structure constants depend on
normalization choices that, however, do not affect
the crosscap constraint.  We use a somewhat
non-standard convention, so that the two-dimensional two-point functions on the
sphere
are normalized to the quantum dimensions of the fields, and therefore
$C_{II0}=[ 2 I + 1 ]$.
This is a natural choice when one exhibits the internal quantum group
symmetry of the model. A determination of the signs
$\epsilon_I$
for the minimal models may be found in \cite{pz} and references therein.

The crosscap constraint is a vastly overdetermined system, and therefore here
we shall only
exhibit a set of equations sufficient to determine all the non-vanishing
$\Gamma_{l}$.  Let us therefore
concentrate on the case $k=4 \rho + 2$, relevant to the non-diagonal
$D_{2 \rho + 3}$ models.
{}From the amplitudes for two fields of isospin
$I_1 = {\bar I}_1 = I$, $I_2 = {\bar I}_2 = 1, \ (I=1,..,k/2-1)$, present both
in the $A$ and in the $D_{odd}$ models, one finds
\be
\Gamma_{I+1} \ \sqrt{ [ 2 I + 3 ]} = \Gamma_I \ {{[2]} \over {[2 I]}} \
\sqrt{ [ 2 I + 1 ]}
\ + \ {{\Gamma_{I-1}} \over {\sqrt{ [ 2 I - 1 ]}}} \left( [2 I + 1] -
{{[2]} \over {[2 I]}} \right) \quad . \label{recursion1}
\ee
On the other hand, from the amplitude for two fields of isospins
$I={\bar I}=k/4$ (hence half-odd-integer), one finds
\be
\sum_{l=0}^{k/2} \ (-1)^l \ \Gamma_l \ \sqrt{[ 2 l + 1 ]} \ + \
\varepsilon_{({k \over 4},
{k \over 4})} \
\Gamma_0 \ \left[ {k/2} + 1 \right] \ = \ 0 \label{recursion2d}
\ee
for the $A$ model, and
\be
\sum_{l=0}^{k/2} \ \Gamma_l \ \sqrt{[ 2 l + 1 ]} \ + \ \varepsilon_{({k \over
4},
{k \over 4})} \
\Gamma_0 \ \left[ {k/2} + 1 \right] \ = \ 0 \label{recursion2nd}
\ee
for the $D_{odd}$ model, where $\varepsilon$ is the same phase that enters
eq. (\ref{cconstraint}).  These equations allow one to determine
recursively all the coefficients $\Gamma_{l}$
up to a common normalization factor that can be fixed by comparison with the
Klein-bottle amplitude.  The final result can be presented in a rather
compact form in terms
of the modular matrices $S$ and $P$.  For the real $D_{2 \rho + 3}$
model with $\varepsilon_{(I,I)}=1$ for all fields,
\be
\Gamma_a = {(-1)^{a^2-1 \over 8} \ P_{{k \over 2},a} \over \sqrt{S_{1,a}}}
\quad ,
\label{dreal}
\ee
while for the complex $D_{2 \rho + 3}$ model with $\varepsilon_{(I,I)}=-1$ for
all half-odd-integer
isospin fields,
\be
\Gamma_a = {(-1)^{a^2-1 \over 8} \ P_{{k \over 2}+2,a} \over \sqrt{S_{1,a}}}
\quad ,
\label{dcomplex}
\ee
where, as in the previous Section,
the weight $a$ is related to the isospin $I$ of the characters by $a=2I+1$.
Since both $P_{{k \over 2},a}$ and $P_{{k \over 2}+2,a}$ vanish
for even $a$ \cite{pss}, the phase factors in eqs. (\ref{dreal})
and (\ref{dcomplex}) are just
signs.  One can verify that these solutions satisfy all of eqs.
(\ref{cconstraint}).
\vskip 24pt
\begin{flushleft}
{\large \bf Descendants of the $D_{odd}$ Models}
\end{flushleft}

In this Section we construct the open descendants of the $D_{odd}$ models.  The
starting point for
the $D_{2 \rho + 3}$ model (with level $k = 4 \rho + 2$) is the torus partition
function \cite{ciz}
\be
T \ = \ \sum_{odd \ a=1}^{k+1} \ |\chi_{a}|^{2} \ + \
\sum_{even \ a=2}^{k/2-1} \ ( \chi_{a} \bar{\chi}_{k+2-a} + h.c. ) \ + \
 |\chi_{k/2+1}|^{2} \quad . \label{tdodd}
\ee
The two sets of solutions of the crosscap constraint obtained in the previous
Section imply
that there are two different vacuum Klein-bottle amplitudes.
Anticipating the real (complex) structure of the Chan-Paton charge spaces for
these
models, we shall denote the two solutions corresponding to eqs. (\ref{dreal})
and (\ref{dcomplex})
by ${\tilde K}^r$ and ${\tilde K}^c$ respectively.
These determine the two Klein-bottle projections,
\be
K^r \ = \ {1 \over 2} \left( \sum_{odd \ a=1}^{k+1}  \chi_{a} \ - \
\chi_{k/2+1} \right)  \label{kreal}
\quad
{\rm and}
\quad
K^c \ = \ {1 \over 2} \left( \sum_{odd \ a=1}^{k+1}  \chi_{a} \ + \
\chi_{k/2+1} \right) \quad . \label{kcomplex}
\ee

As already stressed, the correspondence between the signs in $K$ and the phases
in eq.
(\ref{phase}) is $\varepsilon_{(I,I)} = (-1)^{2I} K_{2I+1}$  where, as
in eq. (\ref{klein}), $K_a$ is the coefficient
of $\chi_a$ in the Klein-bottle projection.  In addition, for the fields that
do not enter the
Klein-bottle projection, $\varepsilon$ is determined by the sign in the
corresponding diagonal model. As for the $D_5$ model, one can determine the
annulus and M{\"o}bius partition functions by turning on one charge at a
time.  The resulting maximal dimensions of
the Chan-Paton charge spaces are equal to $\rho +3$ both for the real
and for the complex model with
$k=4 \rho +2$.  This should be contrasted with the corresponding dimension for
the diagonal
models \cite{pss}, equal to $k+1$. In addition, the complex model contains only
one pair of
complex charges.  Remarkably,
as in the diagonal models, both the annulus and the M{\"o}bius partition
functions
may be expressed in terms of $k+1$ (now linearly dependent) charges, of the
fusion-rule coefficients
\be
{N_{ab}}^c \ = \ \sum_d \ {S_{ad} S_{bd} {S^{\dagger}}_{cd} \over S_{1d}}
\quad , \label{verlinde}
\ee
and of the integer-valued tensor $Y$
\be
{Y_{ab}}^c \ = \ \sum_d \ {S_{ad} P_{bd} {P^{\dagger}}_{cd} \over S_{1d}}
\quad . \label{yassen}
\ee
More precisely, for the real model
\ba
A^r \ &=& \ {1 \over 2} \ \sum_{a,b,c} \ \chi_a \ {N_{bc}}^{k+2-a} \ n_b \ n_c
\label{adoddr} \quad , \\
{\tilde A}^r \ &=& \ {1 \over 2} \ \sum_{a} \ \chi_a \ (-1)^{a-1} \
\left( {\sum_{b} \ S_{ab} \ n_b \over \sqrt{S_{1a}}} \right)^2 \label{atoddr}
\quad ,  \\
M^r \ &=& \ \pm \ {1 \over 2} \ \sum_{a,b} \ {\hat{\chi}}_a \ {Y_{b{k \over
2}}}^{a} \ n_b
\label{mdoddr}  \quad , \\
{\tilde M}^r \ &=& \ \pm \ {1 \over 2} \ \sum_{a,b} \ {\hat{\chi}}_a \
\left( {P_{{k \over 2}a} \ S_{ab}  \over S_{1a}} \right) n_b  \quad .
\label{mtoddr}
\ea
In proving eqs. (\ref{adoddr}) and (\ref{mdoddr}), one needs the relation
\be
{N_{bc}}^{k+2-a} \ = \ \sum_d \ (-1)^{d-1} \ {{S^{\dagger}}_{ad} \
S_{bd} \ S_{cd} \over S_{1d}} \quad , \label{verlinde2}
\ee
implied by
\be
\sum_b \ S_{ab} (-1)^{b-1} S_{bc} \ = \ \delta_{a+c,k+2} \quad .
\label{identity}
\ee
The $k+1 \ (=4 \rho +3)$ charges $n$ may then be expressed in terms of
$\rho +3$ independent charges $l$ as
\be
n_a \ = \ n^{\prime}_a \ + \ {i \over {2 \sqrt{\rho +1}}} \
O_a (-1)^{a-1 \over 2} \
(l_{\rho +2} - l_{\rho +3} ) \quad , \label{reduction1}
\ee
where $O_a$ denotes the projector on odd $a$. In addition, the $n^{\prime}$
satisfy the
relations
\ba
& &n^{\prime}_{{k+2 \over 2}+a} \ = \ n^{\prime}_{{k+2 \over 2}-a} \quad
, \qquad
n^{\prime}_{{k+2 \over 4}+a} \ = \ - \ n^{\prime}_{{k+2 \over 4}-a} \qquad ( a
\ge 1 )
\quad , \nonumber \\
& &n^{\prime}_a \ = \ - \ {l_{a+1} \over 2} \quad ,  \qquad
( 1 \le a \le \rho )
\quad ,  \label{reduction2} \\
& &n^{\prime}_{k+2 \over 2} \ = \ l_1 \quad , \qquad
n^{\prime}_{k+2 \over 4} \ = \ {1 \over 2}
\left( l_{\rho +2} + l_{\rho +3} \right) \quad .
\nonumber
\ea

The open sector of the complex model is described by
\ba
A^c \ &=& \ {1 \over 2} \ \sum_{a,b,c} \ \chi_a \ {N_{bc}}^{a} \ n_b
\ n_c \label{adoddc} \\
{\tilde A}^c \ &=& \ {1 \over 2} \ \sum_{a} \ \chi_a \
\left( {\sum_{b \ } S_{ab} \ n_b \over \sqrt{S_{1a}}} \right)^2
\label{atoddc} \\
M^c \ &=& \ \pm \ {1 \over 2} \ \sum_{a,b} \ {\hat{\chi}}_a \ {Y_{{b,{k \over
2}+2}}}^{a} \ n_b
\label{mdoddc} \\
{\tilde M}^c \ &=& \ \pm \ {1 \over 2} \ \sum_{a,b} \ {\hat{\chi}}_a \
\left( {P_{{k \over 2}+2,a} \ S_{ab}  \over S_{1a}} \right) n_b \quad ,
\label{mtoddc}
\ea
with the same charge reduction of eq. (\ref{reduction2}) and, as usual, with
the values of
the two complex charges identified according to
$l_{\rho +3}={\bar l}_{\rho +2}$.
Remarkably, all the coefficients of the $l$ charges in the partition functions
of eqs.
(\ref{adoddr}), (\ref{mdoddr}), (\ref{adoddc}) and (\ref{mdoddc}) are again
integer and
satisfy the general consistency condition for the M{\"o}bius projection given
in eq.
(\ref{constraints}). This presentation of the partition functions has the
virtue of making the
consistency with the fusion algebra
manifest, whereas in the presentation in terms of independent charge sectors
this crucial
property of the construction is somewhat obscured.  These expressions deserve
two additional
remarks.  First of all, the involution
\be
{N_{bc}}^a \leftrightarrow {N_{bc}}^{k+2-a} \qquad {\rm and} \qquad
{Y_{bc}}^a \leftrightarrow {Y_{b,k+2-c}}^a \qquad , \label{realvscomplex}
\ee
interchanges the partition functions of the real and the complex model,
and a similar relation holds for the crosscap coefficients $\Gamma$,
\be
\left( {\Gamma^r}_a \right)^2 \ = \ \left( {\Gamma^c}_{k+2-a} \right)^2 \qquad
{}.
\ee
Moreover, when expressed in terms of the charges $n$, the annulus partition
functions of all
real(complex) off-diagonal models coincide with the partition functions of the
corresponding
complex(real) diagonal ones. The origin of both these correspondences is still
to be better
understood.
\vskip 24pt
\begin{flushleft}
{\large \bf Descendants of the $E_7$ Model}
\end{flushleft}

We now proceed to describe the open and unoriented sectors of the
$E_7$ model. It has level $k=16$ and torus partition function
\be
T_{E_7} \ = \ |\chi_{1}|^{2} + |\chi_{3}|^{2}
+  |\chi_{4}|^{2} +  |\chi_{5}|^{2} +
 \bigl( {\chi}_{6} \bar{ \chi}_{2} + h.c. \bigr) \quad , \label{te7}
\ee
in terms of the generalized characters corresponding to the extended
symmetry of the $D_{10}$ model, whose partition function in this
notation reads
\be
T_{D_{10}} \ = \ \sum_{i=1}^6 \ | \chi_i |^2  \qquad .
\label{td10}
\ee
The Klein-bottle projection of the $E_7$ model is
\be
K_{E_7} \ = \ {1 \over 2} \ \left( \chi_{1} + \chi_{3}
+  \chi_{4} +  \chi_{5} \right) \qquad , \label{ke7}
\ee
and the coefficients in the corresponding vacuum-channel amplitude $\tilde K$
cannot
be expressed as ratios of single matrix elements of $P$ and $S$, as was the
case for
the $D_{odd}$ models.  Rather, the squares of the crosscap coefficients are
linear
combinations of such terms.  The signs of these coefficients are again
determined by
the crosscap constraint of eq. (\ref{cconstraint}).  As in the $D_{odd}$
models,
both the annulus and the M{\"o}bius partition functions of the descendants of
the $E_7$
model
\ba
A_{E_7} \ &=& \ {1 \over 2} \ \biggl( \chi_1 ( l_1^2 + l_2^2 + l_3^2 + l_4^2 )
\ + \nonumber \\
 &&( \chi_2 + \chi_6 ) ( l_3^2 + l_4^2 + 2 l_1 l_4 + 2 l_2 l_3 + 2 l_2 l_4 +
2 l_3 l_4 ) \ + \nonumber \\ & &\chi_3 ( l_2^2 + l_3^2 +2 l_4^2 +
2 l_1 l_2 + 2 l_1 l_3 +2 l_2 l_3 + 2 l_2 l_4
+ 4 l_3 l_4 ) \ + \nonumber \\
 &&\chi_4 ( l_2^2 + 2 l_3^2 + 2 l_4^2 + 2 l_1 l_3 + 2 l_1 l_4 + 2 l_2
l_3 + 4 l_2 l_4 + 4 l_3 l_4 ) \ + \nonumber \\
 &&\chi_5 ( l_1^2 + l_2^2 + l_3^2 + 2 l_4^2 + 2 l_1 l_2 + 2 l_2 l_3
+ 2 l_3 l_4 ) \biggr) \label{ae7}
\ea
\ba
M_{E_7} \ &=& \ \pm \ {1 \over 2} \ \biggl( {\hat \chi}_1
( l_1 + l_2 + l_3 + l_4 ) \ + \
( {\hat \chi}_6 - {\hat \chi}_2 ) ( l_3 +
l_4 ) \ + \nonumber \\  &&{\hat \chi}_3 ( l_2 + l_3 + 2 l_4 ) \ +  \
{\hat \chi}_4 \ ( - l_2 ) \ + \
{\hat \chi}_5 ( l_1 + l_2 + l_3 ) \biggr) \label{me7}
\ea
can be obtained by charge reduction from the corresponding expressions
for the $D_{10}$ model.  Indeed, if in the annulus amplitude
\be
A_{D_{10}} \ = \ {1 \over 2} \ \sum_{a,b,c} \ \chi_a \
{N_{bc}}^{a} n_b n_c \label{ad10}
\ee
the six charges $n$ are expressed in terms of the four independent charges $l$
according to
\ba
& &n_1 \ = \ {1 \over {3 \sqrt{3}}} \ \left( - 2 l_1 + 4 l_2 - l_4
\right) \quad ,
\quad n_2 \ = \ n_6 \ = \ {1 \over {3 \sqrt{3}}} \
\left( - l_1 - l_2 + 3 l_3 + l_4  \right) \quad , \nonumber \\
& &n_3 \ = \ {1 \over {3 \sqrt{3}}} \ \left( 4 l_1 + l_2 + 2 l_4
\right) \quad ,
\quad n_4 \ = \ {1 \over {3 \sqrt{3}}} \ \left( - l_1 + 2 l_2 + 4 l_4
\right) \quad , \label{e7red} \\
& &n_5 \ = \ {1 \over {3 \sqrt{3}}} \
\left( 2 l_1 + 2 l_2 + 3 l_3 - 2 l_4  \right) \quad , \nonumber
\ea
one recovers the $E_7$ amplitude of eq. (\ref{ae7}). One may also
recover the M{\"o}bius amplitude of eq. (\ref{me7}) provided its transverse
channel is defined in terms of the proper $\Gamma_{a}$.

We have thus completed the classification of the open descendants for
all the $ADE$ $SU(2)$ modular invariants of Cappelli, Itzykson and
Zuber.  In particular, we have shown that all models with non-diagonal
left-right pairing
can be obtained by charge reduction from the
corresponding diagonal ones. One can envisage further
extensions, in particular to $SU(3)$, cosets and supersymmetric models,
with many possible applications to open-string theories.
\vskip 36pt
\begin{flushleft}
{\large \bf Acknowledgments}
\end{flushleft}

The authors are grateful to the ``Centre de Physique Th{\'e}orique'' of the
{\'E}cole Polytechnique for the kind hospitality extended to them while this
work was being completed. This work was supported in part by
E.E.C. Grant CHRX-CT93-0340.

\vfill\eject

\end{document}